\documentclass{article}
\usepackage{frascatiphys}

\newcommand{\lsim}{\mathrel{\rlap{\raisebox{.3ex}{$<$}}
    \raisebox{-.6ex}{$\sim$}}}
\begin{document}
\title{ 
ATMOSPHERIC NEUTRINOS IN 2002
}
\author{
Edward T. Kearns\\
{\em Physics Department, Boston University, Boston, MA 02215, U.S.A.}
}
\maketitle
\baselineskip=11.6pt
\begin{abstract}
In this talk, I will present a snapshot of key results regarding
atmospheric neutrinos, as of the summer of 2002. I will review the
evidence for neutrino oscillation, dominated by the large sample
from Super-Kamiokande, but supported by confirmation from MACRO
and Soudan 2. I will then review some of the more detailed inquiries
made using Super-Kamiokande data such as alternative scenarios,
three-flavor oscillation, and tau neutrino appearance.
\end{abstract}
\baselineskip=14pt
\section{Introduction}

Atmospheric neutrinos are produced by the decay of pions, kaons, and muons in
cosmic ray showers. They have provided a fortuitous laboratory for neutrino
oscillations, where the oscillation probability:
\begin{equation}
P(\nu_a \rightarrow \nu_b) = sin^2 2\theta sin^2 \frac{1.27 \Delta m^2 L}{E}
\end{equation}
is probed over a wide range of $L$ (10-10000 km) and $E$ (0.5-100 GeV).

Two normalizing principles are used to understand the data. First, cosmic ray
showers consist mostly of pions, which decay to $\mu + \nu_\mu$, and the
$\mu$ decays to $e + \nu_\mu + \nu_e$, resulting in a flux ratio
$(\nu_\mu/\nu_e) \sim 2$. This ratio grows to larger values at high energy,
as the time-dilated muon may strike the ground before decaying (removing
$\nu_e$ from the beam), but at all relevant energies the ratio is predicted
to about 5\% accuracy. The second principle is that at equal zenith and nadir
angles the flux of high energy neutrinos must be nearly identical.  Again,
this is predicted with an accuracy of a few percent.


Before I review the experimental data, I should mention some developments in
the neutrino flux calculation\cite{nuflux}, which is a key input to more
detailed understanding of the data. First, there is new data providing
clarity regarding the flux of primary cosmic rays, which is the starting
point for these calculations. The BESS and AMS spectrometers have made new
precision measurements\cite{bess,ams} in the important range of 5-100
GeV/nucleon, with data in good agreement with each other whereas older
measurements had fairly sizeable disagreements.

\begin{figure}[th]
\vspace{4.2cm}
\includegraphics{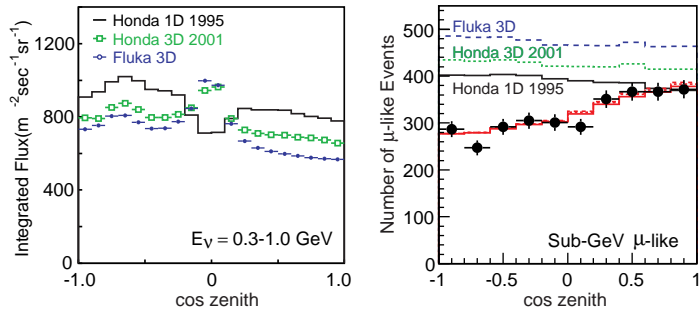}
\caption{\label{fig:3dflux}
  This figure compares 1-dimensional and 3-dimensional flux calculations for
  the Super-Kamiokande site. The left panel compares the integrated $\nu$
  flux from 0.3 to 1 GeV as a function of zenith angle. The right panel shows
  the zenith distribution of $\mu$-like data and Monte Carlo events in the
  Super-K analysis. The lower curves include $\nu$-oscillation.}
\end{figure}

Second, increased computational power and the large Super-K data sample have
motivated the transition from 1-dimensional calculations\cite{1dflux}, where
the $p_t$ of the daughter particles is neglected and every neutrino is
assumed to follow the trajectory of the primary proton, to 3-dimensional
calculations\cite{3dflux}. In 3-dimensions, one should account for $p_t$ as
well as bending in the geomagnetic field. The first 3-D calculations showed
a relatively surprising effect that turns out to play little if any role
in oscillation studies. The effect is an enhanced neutrino flux at the
horizon.  Indeed, below $\sim 1$ GeV, there is fairly sharp peaking as a
function of zenith angle, as shown in Fig.~\ref{fig:3dflux}.  The central dip
in the 1-D curve is caused by geomagnetic cutoff; the central enhancement in
the 3-D curves overwhelms the cutoff. These effects are stronger with
decreasing neutrino energy, but the lower the energy of the neutrino, the
poorer is the correlation between the neutrino direction and that of the
outgoing lepton.  The net result is that the any structure at the horizon is
washed out of the angular distributions of the lepton, as also shown in
Fig.~\ref{fig:3dflux}.  The predicted shapes are nearly the same for each
calculation. There are some differences in the overall normalization between
the models, but in neutrino oscillation studies, the flux normalization is
usually allowed to float as a free parameter.

\section{Evidence for Neutrino Oscillations}

The evidence for neutrino oscillations\cite{sk-evidence} hinges on the two
aforementioned normalizations of the atmospheric neutrino flux: that it
should be composed of a well-predicted ratio of $\nu_\mu$ to $\nu_e$, and
that it should be up-down symmetric at all zenith angles. To further cancel
experimental systematic uncertainties, the flux ratio of $\nu_\mu$ to $\nu_e$
is compared to a detailed Monte Carlo simulation by forming a double-ratio,
$R$. For the sub-GeV sample of events ($E_\nu \lsim 1.5$ GeV), the
Super-Kamiokande analysis of 1489 days of data\cite{sk} shows:
\begin{equation}
R = \frac{(N_\mu/N_e)_{data}}{(N_\mu/N_e)_{m.c.}} = 0.688 \pm 0.016 \pm 0.050.
\end{equation}
The up-down asymmetry is expected to be zero for the multi-GeV muon sample
($E_\nu > \sim 1.5$ GeV), but is measured to be:
\begin{equation}
A = \frac{N_{up} - N_{down}}{N_{up} + N_{down}} = -0.303 \pm 0.030 \pm 0.004,
\end{equation}
whereas the asymmetry is consistent with zero for the multi-GeV electron
sample and the multi-GeV muon Monte Carlo. The up-down asymmetry deviates
from expectation by more than 10$\sigma$, essentially independent of any
Monte Carlo input.

\begin{figure}[th]
\vspace{8.5cm}
\includegraphics{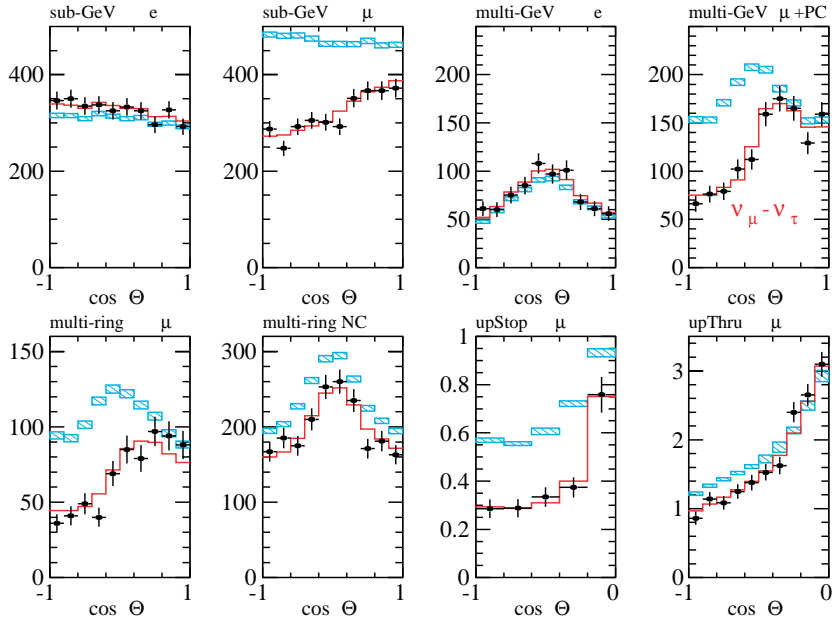}
\caption{\label{fig:rate-zen-all}
The zenith angle distribution for the atmospheric neutrinos used
in the Super-Kamiokande oscillation analysis.}
\end{figure}

\begin{figure}[th]
\vspace{9.0cm}
\includegraphics{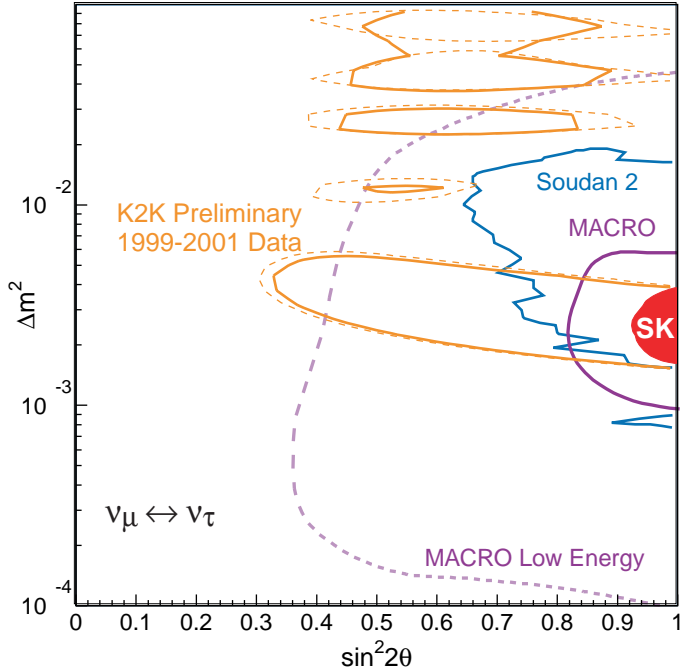}
\caption{\label{fig:contours}
  Confidence intervals for the parameters of atmospheric neutrino
  oscillation, as determined by several different experiments. The two
  similar contours for K2K reflect different treatment
  of systematic errors in the analysis.}
\end{figure}

To estimate the parameters of neutrino oscillation, Super-K bins the
single-ring $\nu_e$ and $\nu_\mu$ events in energy and zenith angle
(effectively path-length $L$). Additional $\nu_\mu$ bins are also used:
partially contained events where the muon exits the detector, multiple-ring
events where the brightest ring is $\mu$-like, or upward-going muons where
the muon enters the detector from a neutrino interaction in the surrounding
rock. Another category of multiple-ring events are selected where the
brightest ring is $e$-like; these have an enhanced fraction of neutral
current ($\simeq 35\%$). These bins play a negligible role in the $\nu_\mu
\leftrightarrow \nu_\tau$ fits but are valuable in restricting alternative
models involving sterile neutrinos or neutrino decay.  In total, 155 bins are
formed, too many to easily display.  Figure~\ref{fig:rate-zen-all} shows 95
bins in 10 zenith angle distributions, where some of the 155 bins have been
combined.  The data fits the prediction beautifully if a two-flavor $\nu_\mu
-\nu_\tau$ oscillation is applied to the neutrino flux with $\sin^2 2\theta =
1$ and $\Delta m^2 = 3 \times 10^{-3}$ eV$^2$. The contours for this best fit
are shown later, in Fig.~\ref{fig:contours}.

In the past, the value of $R$ measured by iron calorimeters such as
NUSEX\cite{nusex} and Frejus\cite{frejus} seemed to show little evidence for
neutrino oscillations, compared to the contemporaneous water Cherenkov
detectors Kamiokande and IMB.  Currently, the Soudan 2 fine-grained iron
tracking calorimeter in Minnesota, U.S.A., has recorded the largest data
set\cite{s2} of contained atmospheric neutrino vertices using iron. They
confirm the picture of atmospheric neutrino oscillation, with some quite
different systematics such as geomagnetic location, target nucleus, and
reconstruction technique. For their analysis of 5.9 kt$\cdot$yrs of data,
using a minimum momentum of 300 MeV/c, Soudan 2 reports $R = 0.71 \pm .09$,
in agreement with the results from water Cherenkov experiments.

\begin{figure}[th]
\vspace{8cm}
\includegraphics{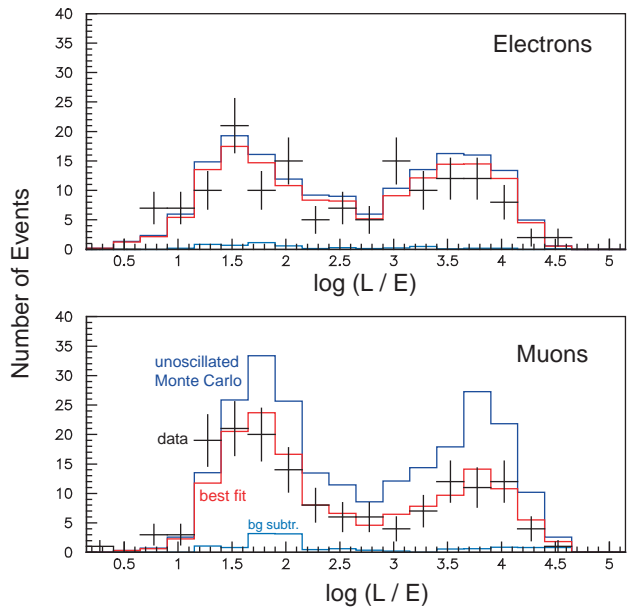}
\caption{\label{fig:soudan2-lovere}
The $L/E$ distribution for high resolution atmospheric neutrinos
in the Soudan 2 analysis.}
\end{figure}

In addition, the Soudan 2 group exploits the good charged particle
recognition of the detector to define a high resolution sample. The sample
includes high energy quasi-elastics (single tracks), low energy
quasi-elastics where the recoil proton from $\nu+n \rightarrow l+p$ is
identified, and high energy multi-prong events. In each case, the neutrino
direction is well determined, with approximately $25^\circ$ angular
resolution (for $\nu_\mu$) and resolution in $\log(L/E)$ of $\sim 0.5$. The
data is binned by $L/E$ and compared to Monte Carlo expectation for different
oscillation parameters.  Figure~\ref{fig:soudan2-lovere} shows the best
agreement, found at $1 \times 10^{-2} {\rm eV^2}$ and $\sin^2 2\theta =
0.97$. The apparent suppression of downward-going $\nu_\mu$
($\log(L/E) < 2.5$) explains the high value of $\Delta m^2$ compared to the
Super-K best fit value. When uncertainties are
incorporated, the 90\% confidence interval for this fit does allow $\Delta
m^2$ down to a few times $10^{-3} {\rm eV^2}$.

\begin{figure}[th]
\vspace{8.0cm}
\includegraphics{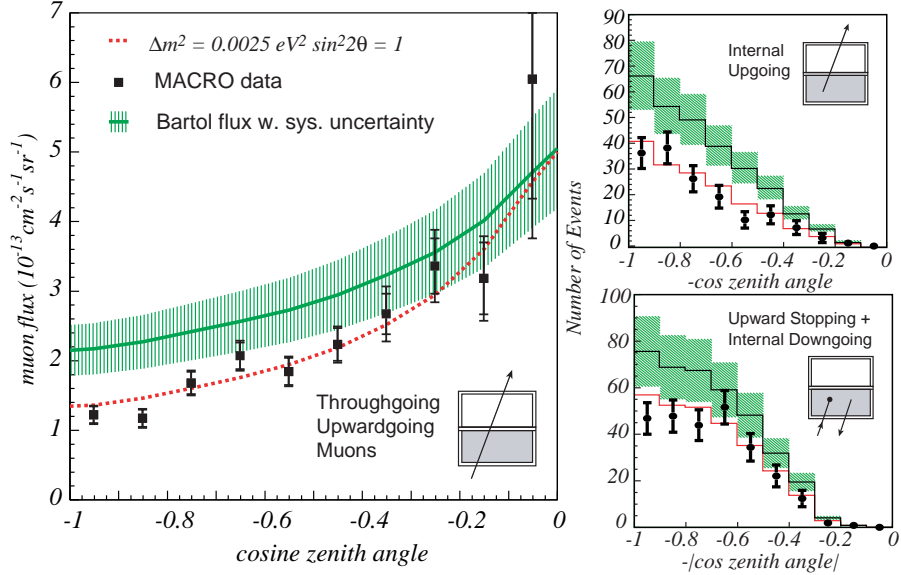}
\caption{\label{fig:macro-data}
The zenith angle distribution for atmospheric neutrinos
of various categories in the MACRO oscillation analysis.}
\end{figure}

The MACRO detector is too coarsely grained and lightweight (4.7~kt) to
effectively study contained interactions.  However, it has a solid angle
acceptance that rivals that of Super-Kamiokande, and is ideally suited to
identify neutrino induced upward-going muons using time-of-flight.  MACRO
consists of a lower section filled with crushed rock absorber and an upper
section that is hollow; both sections are instrumented with planes of
streamer tube tracking.  Fast timing is measured by tanks of liquid
scintillator that surround the detector, and the plane between upper and
lower sections is also a timing plane of liquid scintillator. The simplest
analysis\cite{macro} is to measure the distortion in zenith angle flux for
upward muons that pass completely through the detector; this is shown in the
left panel of Fig.~\ref{fig:macro-data}.  These data demonstrate a best-fit
to neutrino oscillation of $2.5 \times 10^{-3} {\rm eV}^2$ and $\sin^2
2\theta = 1$, in perfect agreement with the Super-K result. MACRO has also
studied the approximate energy distribution of these events (in four bins of
14, 35, 90, and 150 GeV) by considering the amount of multiple
scattering\cite{macro-scat}. This analysis shows consistency with neutrino
oscillations, with the greatest $\nu_\mu$ disappearance in the lowest energy
bins.

MACRO also measures the rate of lower energy ($\sim$ 4 GeV) neutrino events
that do not pass all of the way through the detector\cite{macro-low}.
Internal upgoing events originate in the lower absorber but can be
distinguished from downward stopping muons if they pass through two upper
timing planes. In addition, there is a sample of internal downgoing events
that start in the absorber and exit through the bottom. These have an
indistinguishable topology from neutrino induced upward stopping muons.
Fortunately, both of these topologies are neutrino induced, with similar
parent energy distributions. The results for these samples are shown in the
right panels of Fig.~\ref{fig:macro-data}. These data also favor neutrino
mixing in the range of $10^{-3}$ to $10^{-2} {\rm eV}^2$.

Figure~\ref{fig:contours} shows the confidence intervals for the atmospheric
neutrino analyses just described. The Super-Kamiokande, Soudan 2, and MACRO
results are consistent with each other, with the Super-K result yielding the
most significant and precise estimation of mass splitting and mixing angle.
Despite the varied techniques of these three experiments, they do all rely
upon atmospheric neutrinos, hence a further confirmation using a completely
different neutrino source is desirable. The K2K experiment is the first long
baseline experiment to probe neutrino oscillations with energy and flight
distance comparable to atmospheric neutrinos. K2K uses a 98\% pure $\nu_\mu$
beam with mean energy 1.3 GeV that travels 250 km from KEK to the Super-K
detector. The preliminary results from about one half of the planned running
have recently been made public\cite{k2k}, and the contour from K2K is also
overlayed in Fig.~\ref{fig:contours}; the deepest minimum coincides
exactly with the Super-Kamiokande best fit region.

\section{Unlikely Alternatives}

An important question is whether the Super-K atmospheric neutrino
observations are only consistent with $\nu$-oscillation or could
alternatively be explained by neutrino decay\cite{nudecay}, or other exotic
ideas\cite{exotic} such as violations of Lorentz invariance, flavor changing
neutral currents etc.  Table~\ref{tab:osc-mode} summarizes the results of a
fit to the Super-K data, again binned in by energy and zenith angle for
fully- and partially-contained events, upward going muons, and multi-ring
events. The multi-ring events are subdivided into charged current $\nu_\mu$
and neutral current enhanced samples. The Monte Carlo sample used for
$\nu_\mu \leftrightarrow \nu_\tau$ includes tau lepton appearance.

In Table~\ref{tab:osc-mode} both the absolute $\chi^2$ and the differences in
$\chi^2$ compared to the best fit of $\nu_\mu \leftrightarrow \nu_\tau$
oscillation are reported.  In some cases, the functional form of the
transformation probability approaches that of the accepted hypothesis of pure
$\nu_\mu \leftrightarrow \nu_\tau$. But in each case, there are typically
several bins that constrain the alternative hypothesis.  For example, matter
effects suppress the oscillation probability for high energy $\mu$-like bins
($>5$ GeV $\times [\Delta m^2$ in units of $10^{-3} {\rm eV}^2]$).  For
another example, neutral current events should disappear with sterile
neutrinos or neutrino decay. By observing the differences in $\chi^2$, one
effectively ignores whichever bins are not sensitive to the differences in
model. None of these pure alternatives fit the data very well; the
decoherence model\cite{deco} is the most difficult to reject. Observation of
an oscillatory minimum that is not washed out by $L/E$ resolution would
greatly assist in discrimination.

\begin{table}[t]
\centering
\caption{ \it Alternative fits to Super-Kamiokande atmospheric neutrino data.
}
\vskip 0.1 in
\begin{tabular}{|l|l|c|c|r|} \hline
Mode & Functional Form & $\chi^2$ & $\Delta \chi^2$ & $\sigma$ \\
\hline
\hline
$\nu_\mu \leftrightarrow \nu_\tau$ &
$\sin^2 2\theta sin^2(1.27 \Delta m^2 L/E)$ &
173.8 & 0.0 & 0.0 \\
$\nu_\mu \leftrightarrow \nu_e$ &
$\sin^2 2\theta sin^2(1.27 \Delta m^2 L/E)$ &
284.3 & 110.5 & 10.5$\sigma$ \\
$\nu_\mu \leftrightarrow \nu_{sterile}$ &
$\sin^2 2\theta sin^2(1.27 \Delta m^2 L/E)$ &
222.7 & 48.9 & 7.0$\sigma$ \\
$L \times E$ &
$\sin^2 2\theta sin^2(\alpha L E)$ &
281.6 & 107.8 & 10.4$\sigma$ \\
$\nu_\mu$ decay &
$\sin^4 \theta + \cos^4 \theta e^{-\alpha L/E}$ &
279.4 & 105.6 & 10.3$\sigma$ \\
$\nu_\mu$ decay &
$\sin^2 \theta + \cos^2 \theta e^{-\alpha L/2E}$ &
194.0 & 20.2 & 4.5$\sigma$ \\
$\nu_\mu$ decoherence &
$\frac{1}{2}\sin^2 2\theta(1-e^{-\gamma L/E^2})$ &
184.3 & 10.5 & 3.2$\sigma$ \\
No oscillation &
- &
427.4 & 252.4 & 15.9$\sigma$ \\

\hline
\end{tabular}
\label{tab:osc-mode}
\end{table}

\begin{figure}[th]
\vspace{6.8cm}
\includegraphics{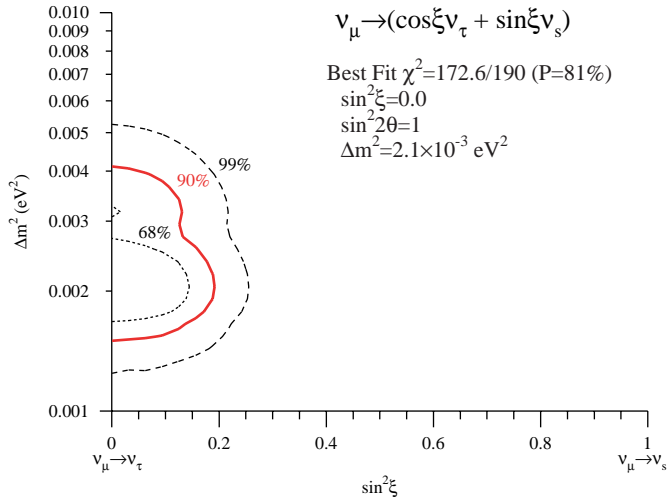}
\caption{\label{fig:sterile-limit}
  The allowed admixture of $\nu_{sterile}$ in a $4\nu$ analysis of
  the Super-Kamiokande data.}
\end{figure}

Clearly, one can generally include any exotic alternative as an admixture
with ``standard'' neutrino oscillations. In fact, one is forced to consider
both in the decay scenarios, where the neutrinos are required to have mass.
For the important case of an admixture of sterile neutrino oscillation,
Super-K has quoted limits on the allowable fraction of sterile neutrinos in a
4-neutrino scheme\cite{fogli-nusterile}. The $\nu_{sterile}$ mixture is
found to be less than 20\% (at 90\% CL), with best fit to the Super-K data is
at pure $\nu_\mu \leftrightarrow \nu_\tau$ mixing.
Figure~\ref{fig:sterile-limit} shows the allowed region a function of the
admixture fraction.

\section{Three Flavor Oscillations}

Having established nearly maximal $\nu_\mu \leftrightarrow \nu_\tau$ mixing
as the explanation for the disappearance of atmospheric neutrinos, it is
logical to search for a smaller component of mixing with $\nu_e$. This is
properly done in a 3-neutrino framework, where a $3 \times 3$ unitary matrix
mixes the mass eigenstates with the flavor eigenstates. Because the earth is
full of electrons, but devoid of free muons or tauons, a component of
electron neutrino flavor passing through the earth may be enhanced by a
matter effect resonance. The size of the resonant enhancement depends on
neutrino energy and the pathlength through matter; therefore, this can
potentially be observed by measuring a larger rate than expected in some
narrow regions of zenith angle and energy. If one mass splitting dominates in
scale over the other, as seems to be the case considering the likely results
for solar neutrinos, the oscillation probabilities $P(\nu_e \leftrightarrow
\nu_\mu)$, $P(\nu_\mu \leftrightarrow \nu_\tau)$, and $P(\nu_\tau
\leftrightarrow \nu_e)$ may be expressed as functions of three parameters:
two mixing angles ($\theta_{13}$, $\theta_{23}$) and the larger mass
splitting $\Delta m^2$.

\begin{figure}[th]
\vspace{5.2cm}
\includegraphics{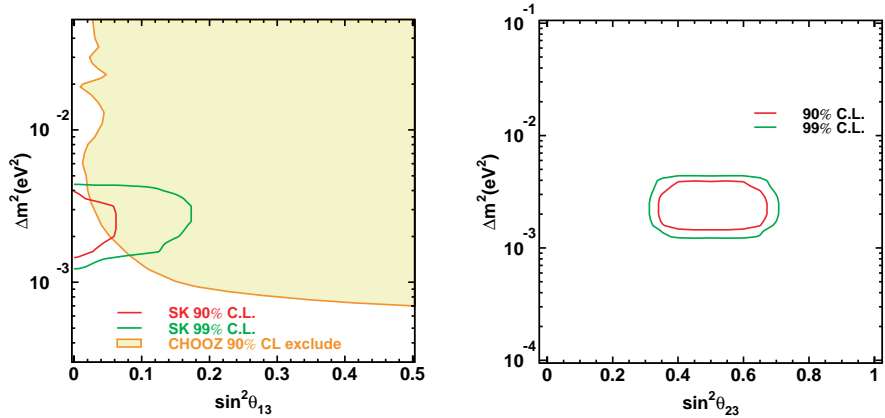}
\caption{\label{fig:3flav-limit}
  Contours for the mixing parameters of three active neutrinos, in
  the single dominant $\Delta m^2$ scenario, as allowed by 1489 days of
  Super-Kamiokande data.}
\end{figure}

The Super-Kamiokande group has analyzed the sample of electron neutrino
interactions with fine binning, but sees no deviation from the predicted
rates. This null result is then turned into limits on the three mixing
parameters, as shown by contours in Fig.~\ref{fig:3flav-limit}.  The left
panel shows that the small fraction of $\nu_e$ appearance allowed by the data
is consistent with the limit on $\nu_e$ disappearance from the
Chooz\cite{chooz} reactor experiment. In fact, the data is consistent with no
$\nu_e$ mixing whatsoever.

\section{Tau Appearance}

If $\nu_\mu \leftrightarrow \nu_\tau$ oscillation is taking place, tau
leptons should appear in the atmospheric neutrino flux. Most atmospheric
$\nu_\tau$ do not have more than the $\approx 3.5 GeV$ of energy needed to
produce a $\tau$ lepton, however, in the large Super-K sample one does expect
approximately 85 charged current $\nu_\tau$ interactions.  The events have
many final state particles above Cherenkov threshold making exclusive
reconstruction difficult, but a statistical analysis is possible. One
preliminary analysis uses a neural network to select for characteristics of
high $p_t$ pion production, based on event variables such as number of rings,
number of ring seeds, number-of-decay electrons, particle classification of
the brightest ring, and so on.

\begin{figure}[th]
\vspace{7.2cm}
\includegraphics{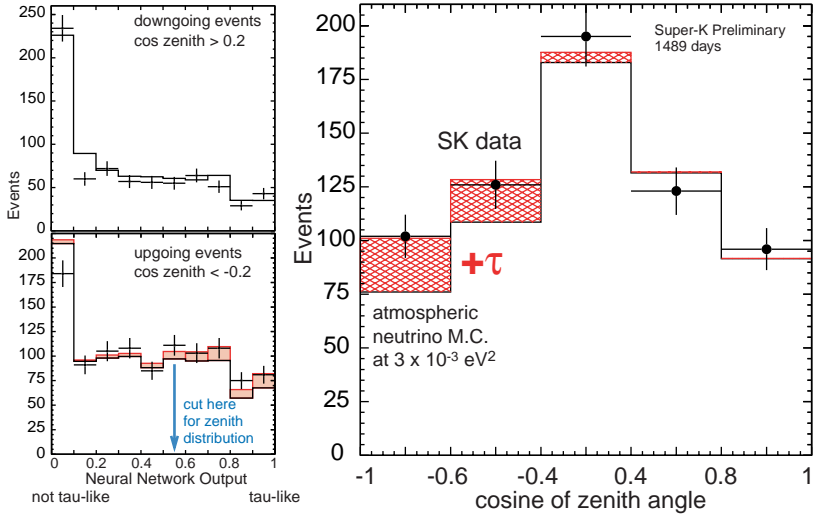}
\caption{\label{fig:tau-result}
  The left panels show the performance of the neural network identification.
  The right panel shows the zenith distribution of events selected as
  $\tau$-like; the filled region shows the relative proportion of $\tau$
  appearance that must be added to the atmospheric neutrino Monte Carlo to
  agree with the data.}
\end{figure}

For $\Delta m^2$ less than a few times $10^{-3} {\rm eV}^2$, it is expected
that charged current tau appearance will only occur in the upward-going
zenith bins. Therefore, the analysis is based on the up-down ratio of a
sample selected for enhanced tau production, and the downward-going data may
be used to check the performance of the neural network, as shown on the left
of Fig.~\ref{fig:tau-result}. This analysis has a 55\% efficiency to save
$\nu_\tau$ interactions in the fiducial volume while keeping 4.5\% of the
background. After cuts, a fit is done to the zenith angle distribution with
components for both the background and the expected tau appearance signal.
The data and fit results are shown on the right of Fig.~\ref{fig:tau-result}.
The best fit implies that the number of charged current $\nu_\tau$
interactions was: $99 \pm 39_{stat} \pm 13_{\Delta m^2} {}^{+0}_{-15(3\nu)}$.
The first systematic error comes from varying the assumed $\Delta m^2$
between 1 and 5 $\times 10^{-3} eV^2$ and the second comes from allowing
three flavor oscillation at the CHOOZ limit.  Although this is only a
2.3$\sigma$ result, it shows that the Super-K data set is consistent with the
expected rate of tau appearance.

\section{Super-Kamiokande Accident and Reconstruction}

On Nov 12, 2001 at 11:01 JST, a single photomultiplier on the bottom of the
Super-Kamiokande tank imploded. At that time, the tank was 2/3 full of water;
it was refilling after PMT replacement work in the preceding summer.  The
imploding tube caused a chain reaction that destroyed 6777 inner PMTs and
1100 outer PMTs. Subsequently, the collaboration made several studies to
determine the mechanism of the chain reaction. A corrective measure was
devised: the PMTs should be encased in a secondary shell consisting of a
transparent acrylic hemisphere over the photocathode and a fiberglass shell
surrounding the rest of the tube. This enclosure is not intended to withstand
static water pressure; instead, several small holes are placed in the shell.
Then, if a PMT were to implode, water rushes into the resulting vacancy more
slowly and a high pressure shock wave is not developed. This has been tested
several times {\it in situ} up to the full level of the Super-K tank. The
tests show that if the center tube of a grid of $3 \times 3$ tubes is
imploded, the damage is contained within the protective vessel, and the eight
surrounding tubes do not implode -- even if they themselves are unprotected.

It will take several years to fabricate replacement photomultipliers for the
inner detector. In the meantime, the detector is being recommissioned using
the remaining tubes plus some spares to populate every other lattice point of
the original photomultiplier tube array. This will be sufficient for nearly
all research to continue, particularly the K2K long baseline experiment. The
detector should be operational by the end of 2002.

\section{Acknowledgements}

I would like to thank the organizers of the Heavy Quarks and Leptons
Workshop, particularly G. Cataldi. I am also thankful for assistance with the
latest experimental results from M. Shiozawa, M. Goodman, and M. Spurio.
Finally, I gratefully acknowledge support from the U.S. Department of Energy.

\end{document}